\documentclass[fleqn,usenatbib, onecolumn]{mnras}

\usepackage{hyperref}
\hypersetup{pdfstartview={XYZ null null 1.1}}
\usepackage{newtxtext,newtxmath}
\usepackage[T1]{fontenc}
\usepackage{ae,aecompl}
\usepackage{graphicx}	
\usepackage{amsmath}	
\usepackage{amssymb}	
\usepackage{subfigure}

\title[Negative Magnetic Diffusivity $\beta$]{Negative Magnetic Diffusivity $\beta$ replacing $\alpha$ effect in Helical Dynamo}

\author[Kiwan Park]{
Kiwan Park,$^{1}$\thanks{E-mail: pkiwan@gmail}
\\
$^{1}$Booyoung 7
}

\date{Accepted XXX. Received YYY; in original form ZZZ}

\pubyear{2019}

\begin{document}
\label{firstpage}
\pagerange{\pageref{firstpage}--\pageref{lastpage}}
\maketitle

\Large
\begin{abstract}
The $\alpha$ effect is known to be an indispensable energy source of the poloidal magnetic field $B_{pol}$ in the sun or planet. However, the $\alpha$ effect is quenched as the magnetic field grows due to the conservation of magnetic helicity. With these conventional understanding, what indeed generates and sustains the observed $B_{pol}$ remains a mystery. To solve this contradiction between theory and the real nature, we derived a semi-analytic representation of $\alpha$ \& $\beta$ using large scale magnetic helicity $\overline{H}_M$ and energy $\overline{E}_M$. Applying the simulation data to $\alpha$ \& $\beta$, we found that the negative $\beta$ effect is a promising substitution of the quenched $\alpha$ effect. However, since the negative $\beta$ effect contradicts the conventional dynamo theory, we derived the new $\beta$ expression referring to the field structure model. This analytic result with the field relation between velocity `$U$' and magnetic field `$B$' shows that the $\beta$ effect in the helical system is not a fixed one. Rather, it plays a variable and dynamic role in dynamo depending on the interaction between the poloidal velocity field $U_{pol}$ and relative strength of large scale magnetic field $\overline{B}$.

\end{abstract}


\section{Introduction and method}
Magnetic field $B$ and plasma (ionized particles) are ubiquitously observed phenomena in space. Interacting with the ionized particles, $B$ field plays an important role in the evolution of celestial plasma systems. Through electromotive force EMF ($\sim\mathbf{U}\times \mathbf{B}$, U: velocity), turbulent plasma energy is converted into magnetic energy which cascades toward the large scale (large scale dynamo, LSD) or small scale regime (small scale dynamo, SSD). As the magnetic field grows, the constraint of magnetic field on the plasma system becomes stronger. The magnetic field controls the rate of collapse and formation of an accretion disk transporting angular momentum (magneto-rotational instability \citep{1991ApJ...376..214B, 2005MNRAS.362..369M}). Also, the balance between the thermal(kinetic) pressure and electromagnetic pressure decides the stability of plasma system (e.g. sausage, kink, or Kruskal-Schwarzschild instability, see \citet{2003phpl.book.....B}). However, since the mutual interaction between the magnetic field and plasma is a coupled nonlinear phenomenon, a minor change can bring about a unexpected considerable consequence.\\

In this paper, we do not discuss the general dynamo theory or magnetic effect (\cite{2005A&A...439..835B}, and references therein). Instead, we will focus on the mathematical and physical properties of EMF as a helical dynamo generator. We suggest a semi-numerical \& analytical method to find the pseudo tensors $\alpha$ \& $\beta$ that linearize EMF or dynamo with the large scale magnetic field $\overline{B}:\, \langle {\bf U}\times {\bf B}\rangle \sim \alpha \overline{\bf B} -\beta \nabla \times \overline{\bf B}$. Then, we investigate the physical meaning of $\alpha$ \& $\beta$ using a field structure model. Especially, we focus on the property of $\beta$ replacing the quenched $\alpha$ effect.\\

\subsection{Numerical method}
All the magnetized plasma phenomena can be explained with kinetic theory or magnetohydrodynamic (MHD) model. Kinetic theory aims to find the distribution density of particles and can provide detailed information on the system. However, the kinetic model is not suitable to the macroscopic description of the dynamically evolving system. Hence we use the MHD approach on a single fluid point of view to explain the dynamo phenomena in the large scale. MHD equations are derived by taking average moments of Boltzmann's kinetic equation. The equation set is composed of continuity, momentum, and magnetic induction equation as follows:
\begin{eqnarray}
\frac{\partial \rho}{\partial t}&=&-{\bf U} \cdot {\bf \nabla}\rho -\rho {\bf \nabla} \cdot {\bf U},\label{continuity_equation_original}\\
\frac{\partial {\bf U}}{\partial t}&=&-{\bf U} \cdot {\bf \nabla}\mathbf{U}-{\bf \nabla} \mathrm{ln}\, \rho + \frac{1}{\rho}(\nabla\times{\bf B})\times {\bf B}\nonumber\\&&
+\nu\big({\bf \nabla}^2 {\bf U}+\frac{1}{3}{\bf \nabla} {\bf \nabla} \cdot {\bf U}\big)+\textbf{f}_{kin},\label{momentum_equation_original}\\
\frac{\partial \mathbf{B}}{\partial t}&=&\nabla \times \langle \mathbf{U}\times \mathbf{B}\rangle +\eta \nabla^2\mathbf{B}+\textbf{f}_{mag}.
\label{magnetic_induction_equation_original}
\end{eqnarray}
Here, $\rho$, $\nu$, and $\eta$ indicate density, kinematic viscosity, and (molecular) magnetic diffusivity in order. The velocity field $U$ is in the unit of sound speed $c_s$, and the magnetic field is normalized by $(\rho_0\,\mu_0)^{1/2}c_s$, where $\mu_0$ is magnetic permeability in vacuum.\\

The general solution of these coupled differential equation is unknown. In this paper, we will use some approximate theoretical models and numerical data. We use Pencil-code that  solves the MHD equations for the compressible conducting fluid in a periodic box \citep{2001ApJ...550..824B}. We forced the plasma system with the helical or nonhelical turbulent kinetic energy ${\bf f}(x,t)$ which is represented like

\begin{eqnarray}
{\bf f}_k(t)=\frac{i\mathbf{k}(t)\times (\mathbf{k}(t)\times \mathbf{\hat{e}})-\lambda |k(t)|(\mathbf{k}(t)\times \mathbf{\hat{e}})}{k(t)^2\sqrt{1+\lambda^2}\sqrt{1-(\mathbf{k}(t)\cdot \mathbf{e})^2/k(t)^2}}.\nonumber
\label{forcing ampliitude fk}
\end{eqnarray}

Here, `$\mathbf{\hat{e}}$' is an arbitrary unit vector, `and `$\phi(t)$' is a random phase ($|\phi(t)|\leq\pi$), and `$\lambda$' denotes the helicity ratio. For example, if `$\lambda$' is `$\pm1$', $i\mathbf{k}\times \mathbf{f}_k=\pm k\mathbf{f}_k$ (fully helical). Also, if `$\lambda$' is `0', $\mathbf{f}_k$ becomes fully nonhelical.\\ 

\subsection{General Analytic method}
The numerical simulation yields the most detailed result. However, it is very difficult to interpret the data  without some appropriate theory. Dynamo theory as well as other MHD model is supposed to solve all MHD equations in principle. However, in many cases, especially for the incompressible system, momentum equation Eq.~(\ref{momentum_equation_original}) and magnetic induction equation Eq.~(\ref{magnetic_induction_equation_original}) are mainly solved with some closure assumption based on the statistical equilibrium state \citep{1976JFM....77..321P, Arkira2011}. Furthermore, some dynamo theories, where the evolution of magnetic field is the main interest, solve only Eq.~(\ref{magnetic_induction_equation_original}) with the assumption of velocity distribution $\langle U_iU_j\rangle$ \citep{1970JETP...31...87V, 1978mfge.book.....M, 1980opp..bookR....K}. Especially, when the field is helical ($\mathbf{f}\sim\nabla \times \mathbf{f}$), Eq.~(\ref{magnetic_induction_equation_original}) for the large scale magnetic field $\mathbf{\overline{B}}$ can be more simplified with $\alpha$ \& $\beta$ (Brandenburg A., Subramanian K., 2005, Astron. \& Astrophysics, 439, 835; Park K., Blackman E. G., MNRAS, 419, 913;
Park K., Blackman E. G., MNRAS, 423, 2120).
\begin{eqnarray}
\frac{\partial \overline{\mathbf{B}}}{\partial t}&=&\nabla \times \langle \mathbf{u}\times \mathbf{b}\rangle  +\eta \nabla^2\overline{\mathbf{B}},\label{LS_magnetic_induction_equation_raw}\\
&\sim& \nabla \times (\alpha \overline{\mathbf{B}})+(\beta+\eta)\nabla^2\overline{\mathbf{B}}.
\label{LS_magnetic_induction_equation_alpha_beta}
\end{eqnarray}

 
So if $\alpha$ \& $\beta$ are available, the nonlinear dynamo process can be described in an intuitive linear way. However,  like the general MHD solution the exact $\alpha$ \& $\beta$ are not yet known. The first order smoothing approximations of the coefficients using MFT are $\alpha= 1/3 \int^t (\langle \mathbf{j} \cdot \mathbf{b} \rangle-\langle \mathbf{u} \cdot \mathbf{\omega} \rangle)\,d\tau$, $\beta= 1/3 \int^t \langle u^2 \rangle\,d\tau$ \citep{1978mfge.book.....M, 1980opp..bookR....K}. However, since the derivation of these results assume small magnetic Reynolds number $Re_M\,(=ul/\eta)$ or Strouhal number St $(=u\tau/l)$\footnote{`$l$' and `$\tau$' are the characteristic length and time scale}, the validity of these coefficients in space where $Re_M$ and $St$ are huge has been under dispute. Moreover, the possibility of quenching $\alpha$ with the growing helical component in magnetic field ($\langle \mathbf{j} \cdot \mathbf{b} \rangle \rightarrow \langle \mathbf{u} \cdot \mathbf{\omega} \rangle$) makes it difficult to explain how the dynamo process is sustained.\\

 Nonetheless, Eq.~(\ref{LS_magnetic_induction_equation_alpha_beta}) itself is still a valid statistical relation of the second order moment regardless of the astrophysical conditions. It becomes clear if we take the inner product of $\overline{\textbf{B}}$ to $\partial \overline{\textbf{B}} /\partial t$ and use $\langle \overline{\bf J}\cdot \overline{\bf B}\rangle=\langle \overline{\bf A}\cdot \overline{\bf B}\rangle$ for the helical large scale field $k=1$ (see Eq.~(\ref{second_order_moment_replacement})):
\begin{eqnarray}
\overline{\mathbf{B}}\cdot \frac{\partial \overline{\mathbf{B}}}{\partial t}
&\sim& \alpha \langle \overline{\mathbf{A}}\cdot \overline{\mathbf{B}}\rangle-(\beta+\eta)\langle\overline{B}^2\rangle,
\label{Statistical_LS_magnetic_induction_equation_alpha_beta}
\end{eqnarray}

If there is some additional forcing source in the MHD system, its effect is implicitly reflected in $\alpha$ \& $\beta$. So, without exact information on the internal flow in the sun or planet, Eq. (\ref{LS_magnetic_induction_equation_alpha_beta}) can be used for the study of magnetic field evolution. In practice, Eq.~(\ref{LS_magnetic_induction_equation_alpha_beta}) is split into the poloidal component $\overline{{\bf B}}_{pol}(=\nabla \times \overline{{\bf A}})$ and toroidal one $\overline{\bf B}_{tor}$\citep{2014ARA&A..52..251C}:
\begin{eqnarray}
&&\frac{\partial \overline{A}}{\partial t}=(\eta+\beta)\bigg(\nabla^2-\frac{1}{\varpi^2} \bigg)\overline{A}-\frac{{\bf u}_p}{\varpi}\cdot \nabla (\varpi\overline{A})+\alpha \overline{B}_{tor},\label{Solar_poloidal_magnetic_field}\\
&&\frac{\partial \overline{B}_{tor}}{\partial t}=(\eta+\beta)\bigg(\nabla^2-\frac{1}{\varpi^2} \bigg)\overline{B}_{tor}+\frac{1}{\varpi}\frac{\partial(\varpi\overline{B}_{tor})}{\partial r}\frac{\partial (\eta+\beta)}{\partial r}-\varpi{\bf u}_p\cdot \nabla \bigg(\frac{\overline{B}_{tor}}{\varpi}\bigg)\nonumber\\
&&-\overline{B}_{tor}\nabla\cdot{\bf u}_p+\varpi(\nabla \times (\overline{A}\hat{e}_\phi))\cdot\nabla{\bf \Omega}+\nabla\times(\alpha\nabla \times(\overline{A}\hat{e}_\phi)),
\label{Solar_toroidal_magnetic_field}
\end{eqnarray}
where $\varpi$ is $r sin \theta$ and $\Omega$ is the angular velocity.\\ 

As these equations show, the $\alpha$ coefficient is a prerequisite for the sustainable poloidal magnetic field $\overline{\textbf{B}}_{pol}$. The $\alpha$ effect has been considered as a main energy source to generate $B_{pol}$ in Parker's solar dynamo model \citep{1955ApJ...122..293P}. However, whereas Parker's $\alpha$ effect is based on the direct mechanical effect of buoyancy and Coriolis force, the $\alpha$ effect discussed here is originated from the interaction between $U$ and $B$ $(\sim J)$ \citep{2017PhRvD..96h3505P}. On the other hand, Babcock-Leighton's model \citep{1961ApJ...133..572B, 1969ApJ...156....1L} considers the sunspot effect including the buoyancy and internal convective flow as a primary source of $B_{pol}$. Recently, observation data of the Sun’s polar magnetic field were directly applied to the magnetic induction equation to reproduce the periodic  solar cycle. It expects the more complete pattern of solar magnetic field in a short period \citep{2007PhRvL..98m1103C}. But, it is unclear if this kind of approach indeed explains the mechanism of solar dynamo.\\


The external helical or nonhelical kinetic energy applied to the system change the forcing source `$\textbf{f}$' in Eq.~(\ref{momentum_equation_original}), (\ref{magnetic_induction_equation_original}). But, as mentioned, the external effect is reflected in $\alpha$ \& $\beta$ without changing the basic form of the equation. So the general $\alpha$ \& $\beta$ coefficient are valuable to the investigation of exact dynamo process. Regardless of the exact solution, they at least provide some parameterized information on the system. To derive $\alpha$ \& $\beta$, a few rigorous analytical methods such as eddy damped quasi normalized markovianized approximation (EDQNM, \cite{1976JFM....77..321P}) and direct interactive approximation (DIA, \cite{Arkira2011}) were suggested and  applied to Eq.~(\ref{momentum_equation_original}), (\ref{magnetic_induction_equation_original}), and EMF $\langle \textbf{u}\times \textbf{b}\rangle$. They yielded qualitatively the same $\alpha$ \& $\beta$ coefficient as those of mean field theory (MFT) in the level of the first-order approximation. It is a reasonable result because the second order moments $\langle UU \rangle$ or $\langle BB \rangle$,  which commonly appear in the calculation, are replaced by the same statistical relation like below (Park K., 2014, Mon. Not. R. Astron. Soc., 444, 3837, references therein):
\begin{eqnarray}
\langle X_l(k)X_m(-k)\rangle&=&P_{lm}(k)E(k)+\frac{i}{2}\frac{k_n}{k^2}\epsilon_{lmn}H(k),
\label{second_order_moment_replacement}
\end{eqnarray}
Here, $E$ indicates the trace of the moment, i.e., energy density $\langle X^2\rangle/2$, and $P_{lm}(k)$ is a projection operator $\delta_{lm}-k_lk_m/k^2$. The physical meaning of $H(k)$ becomes clear if we apply it to helicity.
\begin{eqnarray}
\langle {\bf u}\cdot \nabla \times {\bf u}\rangle&=&\int (-ik_j)\xi_{ijl}\langle u_i(k)u_l(-k)\rangle\,d{\bf k}\nonumber\\
&=&\int H(k)\,d{\bf k}.
\label{FT_second_order_moment_replacement}
\end{eqnarray}
Mathematically, $\alpha$ \& $\beta$ are the representative tensors related to helicity and energy for the transient state of second order moment. They converge to $zero$ when the system gets saturated.\\



These conventional theories MFT, EDQNM, and DIA explain that $\alpha$ is the source of large scale magnetic energy and becomes quenched with the growing helical magnetic energy (see Fig.~1). And, $\beta$ is thought to be related to the turbulent kinetic energy leading to the diffusion of magnetic energy. However, a careful look of the figures shows that $\overline{\mathbf{B}}$ continues growing even after the $\alpha$ effect is quenched. Moreover, $\beta$, which is supposed to be positive, keeps negative. In fact, $\beta$ looks more correlated to the slowly growing $\overline{\mathbf{B}}$ than $\alpha$.\\

So far, theoretical possibility of negative $\beta$ and some numerical result for the special case have been reported. Moffatt derived the $\alpha$ \& $\beta$ in Lagrangian formation \citep{1974JFM....65....1M}, and Kraichnan rederived the magnetic induction equation in the strongly helical system \citep{1976JFM....75..657K}:
\begin{eqnarray}
\frac{\partial \overline{\mathbf{B}}}{\partial t}&=&\eta\nabla^2\overline{\mathbf{B}}+\tau_2\nabla\times \langle\alpha\nabla\times \alpha\rangle\,\overline{\mathbf{B}}\nonumber\\
&=&(\eta-\tau_2A)\nabla^2\overline{\mathbf{B}}.
\label{Kraichnan_alpha_beta}
\end{eqnarray}
(Here $\eta=\tau_1u^2_0,\, \alpha(\mathbf{x},\, t)=(-)1/3\langle \mathbf{u}\cdot\omega \rangle\tau_1$, $\langle \alpha(\mathbf{x},\,t)\alpha(\mathbf{x}',\,t')\rangle$=$A(x-x')D_2(t-t'),\,A\equiv A(0),\, \tau_2=\int^{\infty}D_2(t)\, dt$.) This result implicitly assume the long-lasting stability of helical field and memory effect ($\sim\tau_2$) in the large eddy. The experimental result for the negative $\beta$ with specific Taylor-Green flow was also reported (\citep{2015ApJ...811..135A}, references therein). However, the equation has a large flaw. Eq.~(\ref{Kraichnan_alpha_beta}) makes $\langle B^2 \rangle$ and $\langle {\bf A}\cdot {\bf B}\rangle$ independently evolving discrete quantities.

\subsection{Semi-Analytic method}
To calculate $\alpha$ \& $\beta$ numerically, so called test field method (TFM) is used \citep{2005AN....326..245S}. The basic idea is simple and straightforward. Repeated simulations with the embedded arbitrary large scale magnetic field $\overline{B}^T$ can produce the data for $\mathbf{u}$ \& $\mathbf{b}$. Then, using the basic relation $\mathbf{\xi}_i=\langle \mathbf{u}\times \mathbf{b}\rangle_i = \alpha_{ij}\overline{B}^T_j+\beta_{ijk}\partial \overline{B}^T_j/\partial x_k$, $\alpha_{ij}$ \& $\beta_{ijk}$ can be found. This method indeed provides detailed information on $\alpha_{ij}$ \& $\beta_{ijk}$. However, there are couple of things to be considered. First, it should be checked if $\alpha$ \& $\beta$ are not affected by the large scale $\overline{B}^T$. Basically, $\alpha$ \& $\beta$ are small scale quantities which are easily constrained by $\overline{B}$. Second, it should be also made clear if the method can be applied to the astrophysical system. Being different from the lab experiment, there are few things we can do to the astrophysical system except the observation and measurement of the data.\\

Instead of applying the artificial $\overline{B}^T$, we can find $\alpha$ \& $\beta$ using the data for large scale magnetic helicity $\overline{H}_M (=\langle{\bf \overline{A}}\cdot {\bf \overline{B}}\rangle)$ and energy $\overline{E}_M(=\langle \overline{B}^2\rangle/2)$. From the coupled equations below
\begin{eqnarray}
\frac{d}{dt}\overline{H}_M&=&4\alpha \overline{E}_M-2(\beta+\eta)\overline{H}_M,\label{Hm1}\\
\frac{d}{dt}\overline{E}_M&=&\alpha \overline{H}_M-2\big(\beta+\eta\big)\overline{E}_M, \label{Em1}
\end{eqnarray}
we can calculate the solution as follows \citep{2019ApJ...872..132P}
\begin{eqnarray}
2\overline{H}_M(t)&=&(2\overline{E}_M(0)+\overline{H}_M(0))e^{2\int^t_0(\alpha-\beta-\eta)d\tau}\nonumber\\
&&-(2\overline{E}_M(0)-\overline{H}_M(0))e^{-2\int^t_0(\alpha+\beta+\eta)d\tau},\label{HmSolutionwithAlphaBeta1}\\
4\overline{E}_M(t)&=&(2\overline{E}_M(0)+\overline{H}_M(0))e^{2\int^t_0(\alpha-\beta-\eta)d\tau}\nonumber\\
&&+(2\overline{E}_M(0)-\overline{H}_M(0))e^{-2\int^t_0(\alpha+\beta+\eta)d\tau}.\label{EmSolutionwithAlphaBeta2}
\end{eqnarray}
(Here, we used $\langle {\bf \overline{A}}\cdot {\bf \overline{B}} \rangle=\langle {\bf \overline{J}}\cdot {\bf \overline{B}} \rangle$ in the large scale. Nonhelical component drops when the average is taken to the second order moment. $\overline{H}_M$ is always smaller than $2\overline{E}_M$, but $\overline{H}_M\rightarrow 2\overline{E}_M$ with time in the helically forced system.)\\

Again, $\alpha$ \& $\beta$ are
\begin{eqnarray}
\alpha(t)&=&\frac{1}{4}\frac{d}{dt}log_e \bigg|\frac{ 2\overline{E}_M(t)+\overline{H}_M(t)}{2\overline{E}_M(t)-\overline{H}_M(t)}\bigg|,\label{alphaSolution3}\\
\beta(t)&=&-\frac{1}{4}\frac{d}{dt}log_e\big| \big(2\overline{E}_M(t)-\overline{H}_M(t) \big)\big( 2\overline{E}_M(t)+\overline{H}_M(t)\big)\big|\nonumber\\
&&-\eta.
\label{betaSolution3}
\end{eqnarray}
In Eq.~(\ref{LS_magnetic_induction_equation_alpha_beta}) there is a tricky sign relation between $\overline{\mathbf{B}}$ and $\alpha$. However, the sign issue becomes clear wihle Eq.~(\ref{Hm1}), (\ref{Em1}) are derived from Eq.~(\ref{LS_magnetic_induction_equation_alpha_beta}). For example, if the system is driven with positive kinetic helicity, $\alpha\,(\sim \langle \mathbf{j} \cdot \mathbf{b} \rangle-\langle \mathbf{u} \cdot \mathbf{\omega} \rangle)$ becomes negative so that the second terms in the right hand side of Eq.~(\ref{HmSolutionwithAlphaBeta1}), (\ref{EmSolutionwithAlphaBeta2}) become dominant. Since $2\overline{E}_M$ is always larger than $\overline{H}_M$, the sign of $\overline{H}_M$ and ${E}_M$ become consistent with the simulation result. Moreover, because of the magnetic helicity conservation, the sign of large scale magnetic helicity and small scale one become opposite. In the system forced with helical kinetic energy, the sign relation can be used to separate the large scale field from the small scale one without ambiguity. In contrast, when the system is forced with helical magnetic energy, the sign of $\alpha$ and magnetic helicity are the same, which is also reflected in Eq.~(\ref{HmSolutionwithAlphaBeta1}), (\ref{EmSolutionwithAlphaBeta2}). \\




\section{Result and Analysis}

\subsection{Numerical result}

\begin{figure*}
\centering{
   \subfigure[$f_h=1$]{
     \includegraphics[width=8.5 cm]{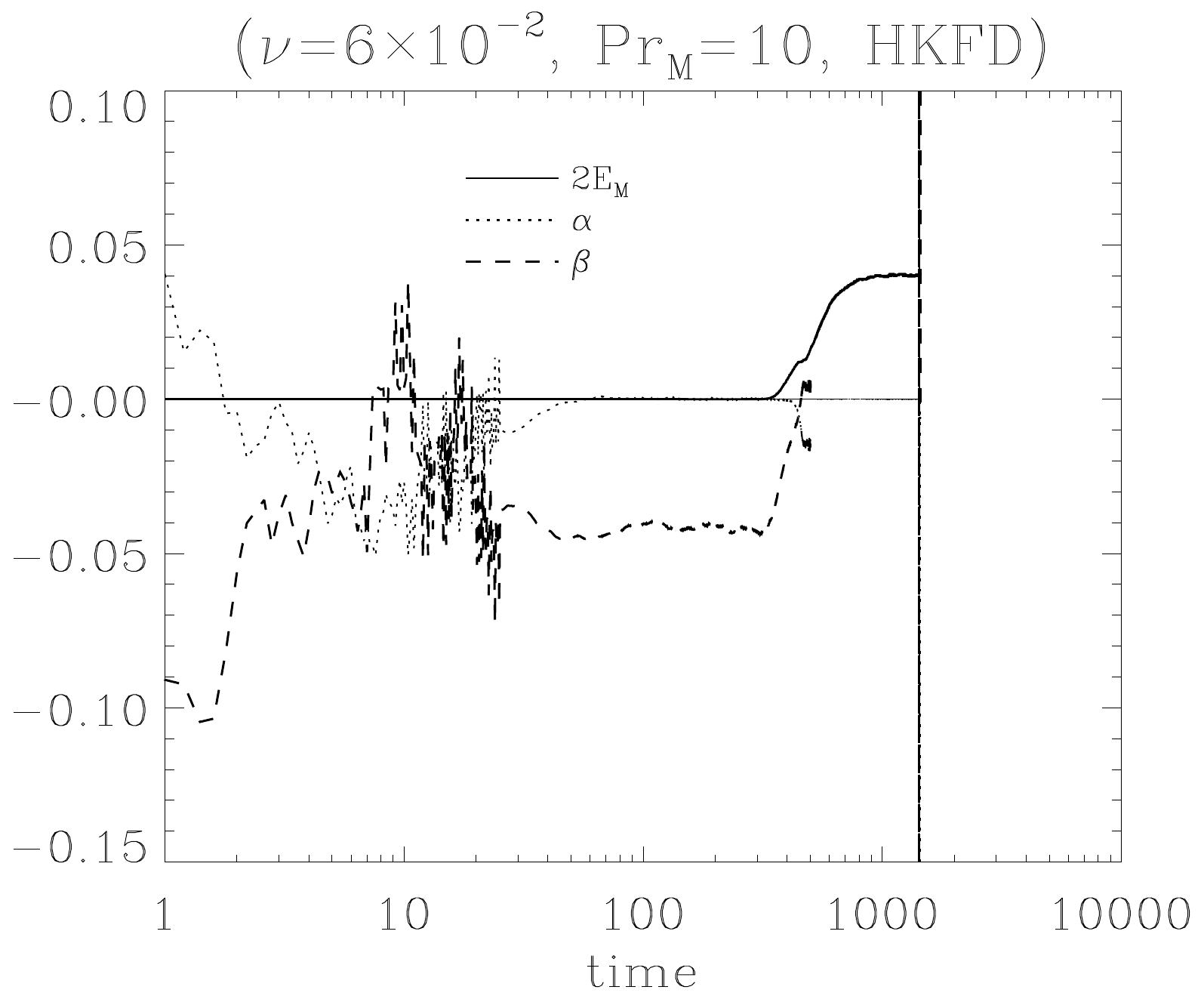}
     \label{f1a}
   }\hspace{0 mm}
   \subfigure[$f_h=1$]{
     \includegraphics[width=8.5 cm]{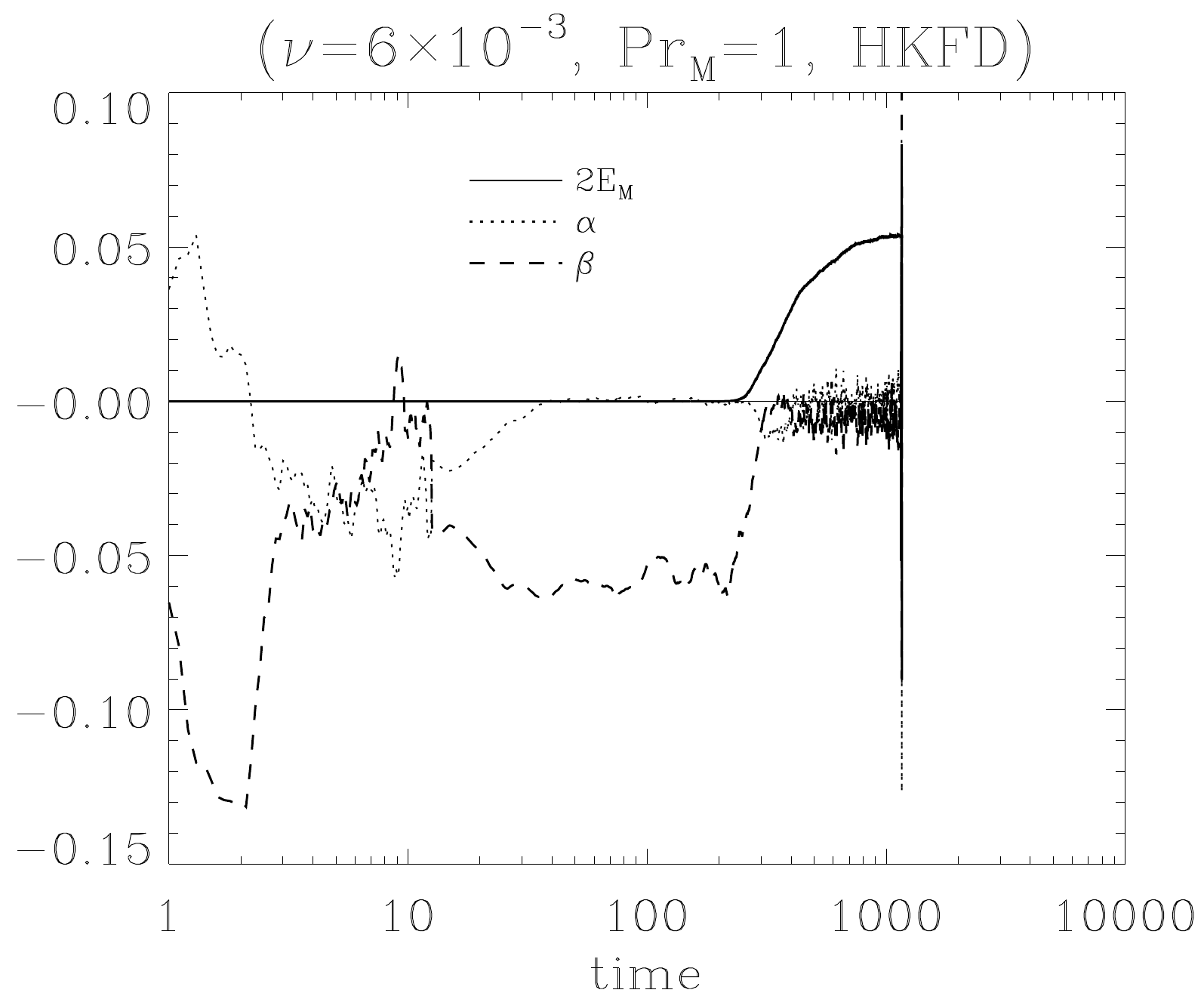}
     \label{f1b}
   }
    \subfigure[$f_h=-1$]{
     \includegraphics[width=8.5 cm]{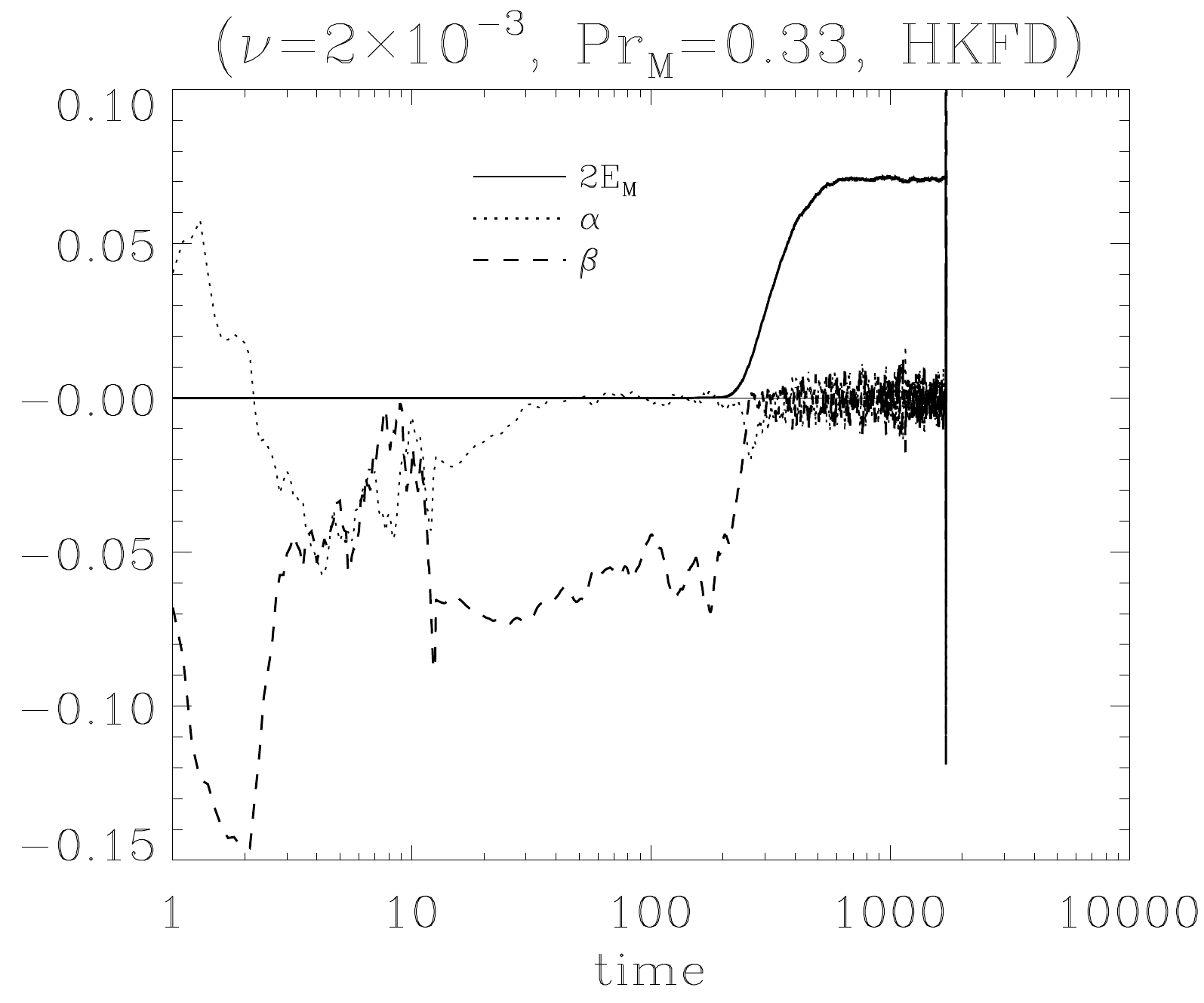}
     \label{f2a}
   }\hspace{-0 mm}
   \subfigure[$f_h=1$]{
     \includegraphics[width=8.5 cm]{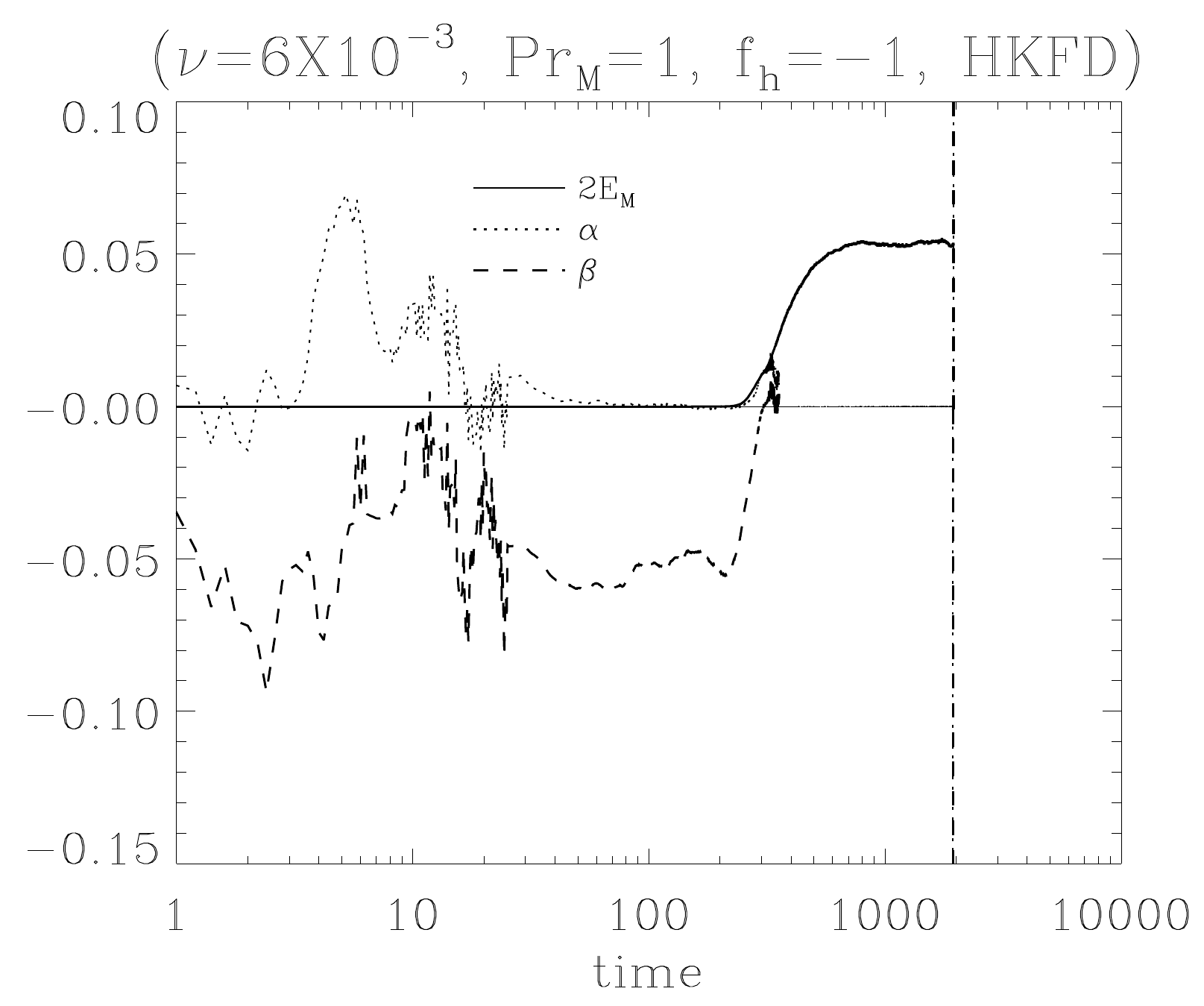}
     \label{f2b}
   }\hspace{-0 mm}
   \subfigure[$f_h=1\rightarrow 0$ at $t\sim 110$]{
     \includegraphics[width=8.5 cm]{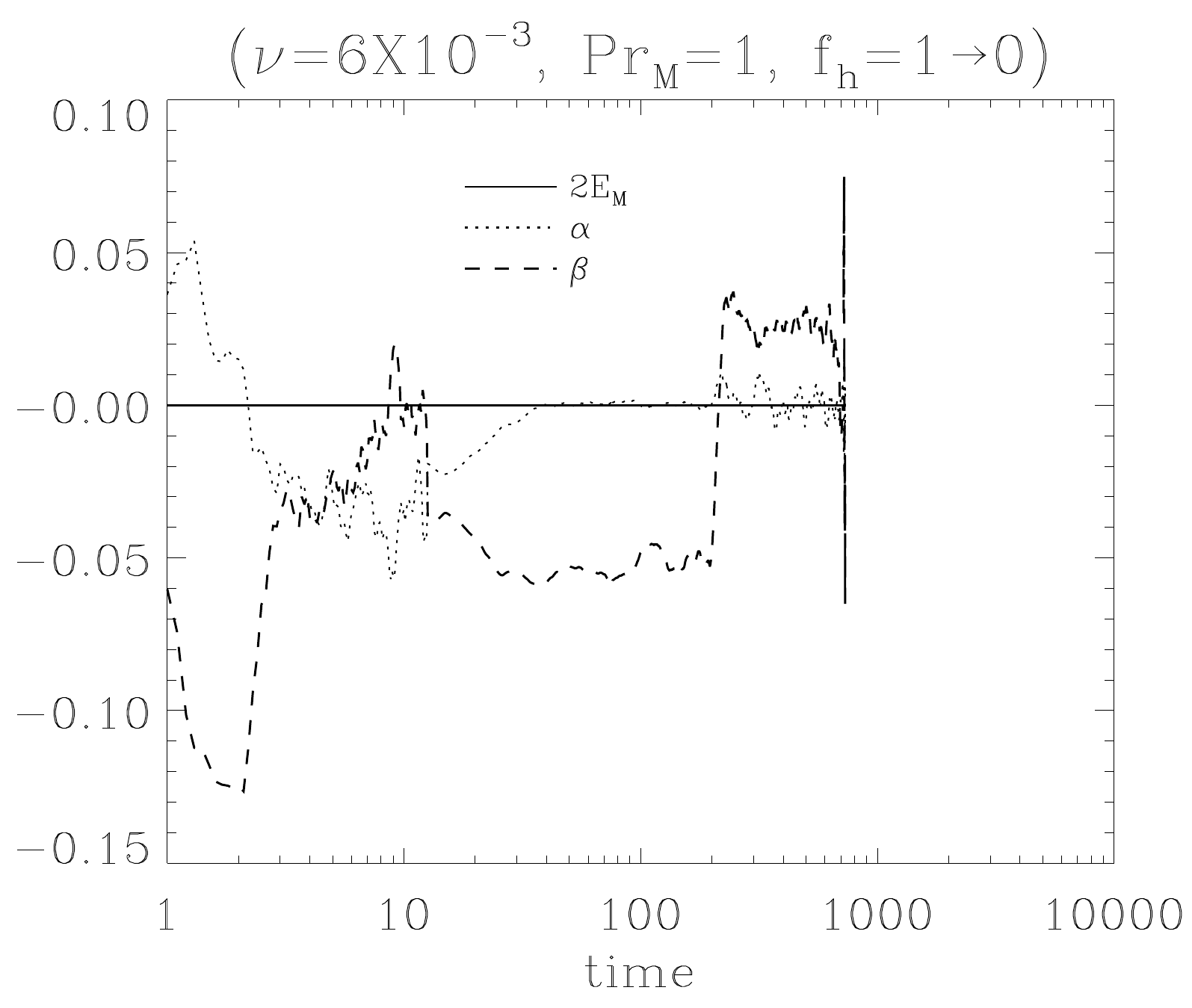}
     \label{f3a}
   }\hspace{-0 mm}
   \subfigure[$f_h=0$]{
     \includegraphics[width=8.5 cm]{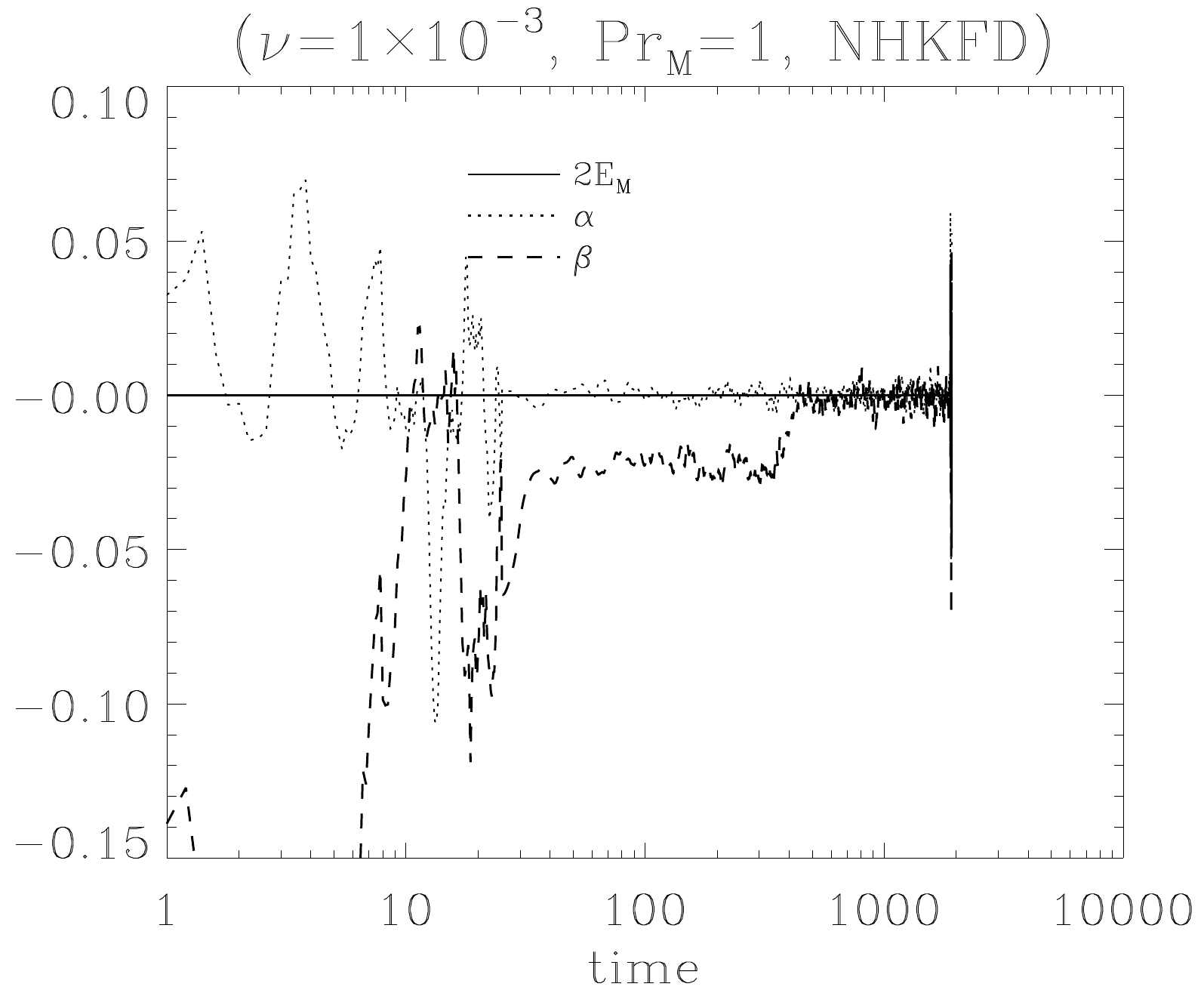}
     \label{f3b}
   }
\caption{We forced six plasma systems with the same kinetic energy but different helicity ratios $f_h \,(=\,\,\langle \mathbf{u}\cdot\omega\rangle /k_f\langle u^2\rangle$ (from $-1$ to 1) and magnetic Prandtl number $Pr_M$ (from 0.33 to 10). Resolution is $400^3$.}
}
\end{figure*}


Fig.~\ref{f1a}, \ref{f1b}, \ref{f2a} show the evolving $\alpha$ (dotted line), $\beta$ (dashed line), and large scale magnetic energy $\overline{E}_M$ (solid line) in the systems forced with the fully right handed helical kinetic energy ($f_h=1$). Their (molecular) magnetic diffusivities $\eta$ are the same, but kinematic viscosities $\nu$ varies to make $Pr_M(=\nu/\eta)$ 10, 1, 0.33 in each system. The evolution of $\alpha$ effect is consistent with the theoretical expectation. However, its quenching position is too early for the slowly growing $\overline{E}_M$. In contrast, $\beta$ keeps some negative value until it becomes quenched with the arising $\overline{E}_M$. The negative $\beta$ is contradictory to the conventional dynamo theory. However, with the negative Laplacian $\nabla^2$ in Fourier space ($-k^2$), it makes sense that the negative $\beta$ effect plays an actual role of forcing the large scale magnetic field (see Eq.~(\ref{LS_magnetic_induction_equation_alpha_beta})).\\

Fig.~\ref{f2b} shows the effect of negative $f_h$ on $\alpha$, $\beta$, and $\overline{E}_M$. The system itself is the same as that of Fig.~\ref{f1b} except the forcing helicity ratio $f_h=-1$. The comparison of these two plots shows that the evolving patterns of $\overline{E}_M$ and $\beta$ are independent of the sign of $f_h$. The $\alpha$ coefficients evolve with the opposite sign, but their magnitudes eventually converge to $zero$. Their temporal profiles are consistent with Eq.~(\ref{HmSolutionwithAlphaBeta1})-(\ref{betaSolution3}).\\

Fig.~\ref{f3a} shows the effect of changing $f_h$. The system starts with the fully helical kinetic energy $f_h=1$. Then, the helicity ratio is dropped to be $zero$ after $\alpha$ is quenched to separate their effects. This sudden change makes $\beta$ elevate up to a positive value. Considering that $\beta$ is independent of the sign of $f_h$ ($+1$ or $-1$), the elevation of $\beta$ is extra ordinary. The positive $\beta$ has the effect of diffusing the magnetic energy, as the conventional dynamo theory expects $\beta\,(\sim u^2)>0$. We will explain its physical mechanism with the analysis of field structure of $U$ \& $B$.\\

Fig.~\ref{f3b} shows the evolution of typical small scale dynamo forced with the fully nonhelical kinetic energy ($f_h=0$). Most of the magnetic energy is cascaded toward the small scale regime, and only partial energy is inversely cascaded to the large scale $\overline{E}_M(\lesssim 10^{-5})$. 
The initial flip-flop $\alpha$ and negative $\beta$ effect in this fully nonhelical forcing system may be responsible for this negligible growth. The nontrivial $\alpha$ effect seems to be caused by the naturally generated magnetic helicity \citep{1958PNAS...44..489W} or some helical component existing in the nonhelical forcing energy in the code. However, the reason is not clear at the moment.\\



\begin{figure*}
\centering{
  {
     \includegraphics[width=19 cm]{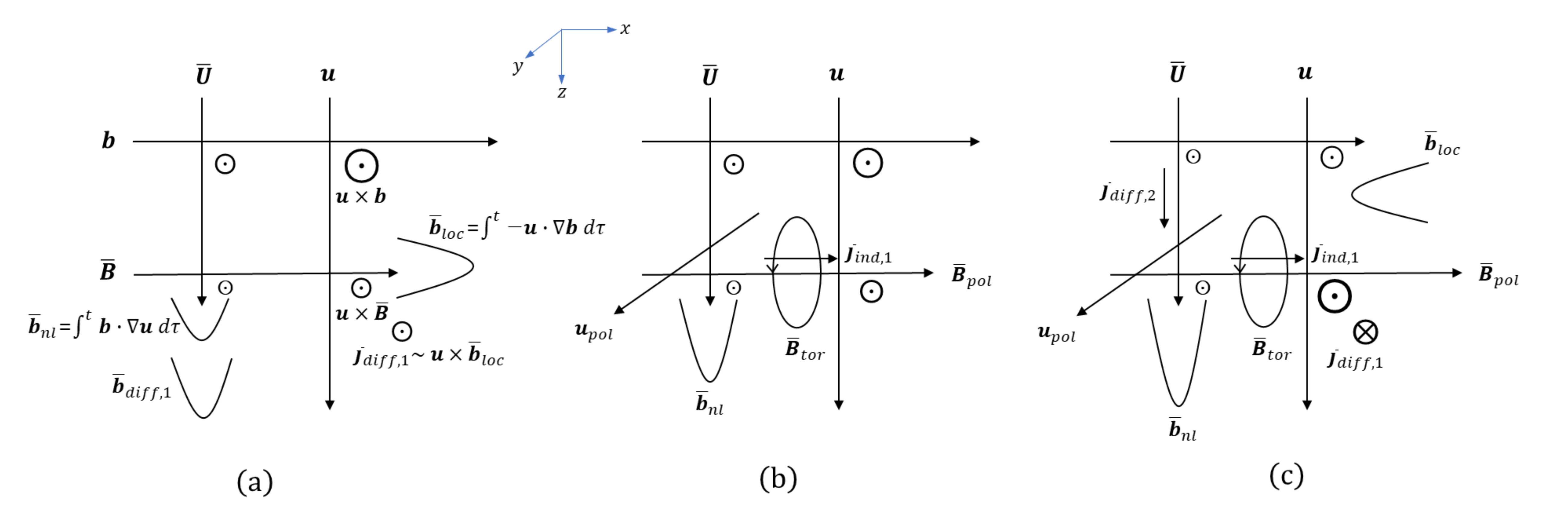}
     \label{f3}
   }
\caption{$\overline{U}$ \& $\overline{B}$ and $u$ \& $b$ represent the large and small scale fields. Between the large and small scale many eddy scales exist. The length of an arrow indicates the strength of a field. In (a) and (b) this structure is based on ${\bf B}\cdot \nabla {\bf U}>0$, $-{\bf U}\cdot\nabla {\bf B}>0$.}
}
\end{figure*}

\subsection{Analysis with field structure}
Eq.~(\ref{HmSolutionwithAlphaBeta1})-(\ref{betaSolution3}) and simulation data explain the temporal evolution of $\alpha$, $\beta$, and $\overline{E}_M(t)$ consistently. From now on, we will discuss the origin of $\alpha$ \& $\beta$ effect and their physical meaning with the evolving large scale magnetic field $\overline{B}$ in helical dynamo. In addition to the semi-analytic equations, we will use the field structure model based on the geometry of $\bf u$ \& $\bf b$ that amplifies $\overline{B}$ \citep{2017MNRAS.472.1628P, 2019ApJ...872..132P}.\\

In Fig.~2(a), we first analyze the geometry of fields structured for the nonhelical system. In this structure,  the geometry of $\bf u$ \& $\bf b$ is constructed to make `$\langle\mathbf{b}\cdot \nabla \mathbf{u}\rangle$' and `$-\langle\mathbf{u}\cdot \nabla \mathbf{b}\rangle$' positive. Their mutual interactions $\langle\mathbf{u}\times \mathbf{b}\rangle_i$ among eddies yield the spatially inhomogeneous current distribution, and their nontrivial curl effect induces(transports) magnetic energy. Magnetic energy at `$b$' converted from plasma energy is locally transferred to `$\overline{B}$' through $\int^t (-\mathbf{u}\cdot\nabla \mathbf{b})\,d\tau\,(\equiv \overline{\mathbf{b}}_{loc})$. Simultaneously, the converted magnetic energy at `$u$' cascades toward $\overline{U}$' through $\int^t (\mathbf{b}\cdot\nabla \mathbf{u})\,d\tau\,(\equiv \overline{\mathbf{b}}_{nl})$.\\

A careful look shows the secondary interaction $\mathbf{u}\times \overline{\mathbf{b}}_{loc}$ is also possible. This is an indirect interaction yielding the current density $\mathbf{j}_{diff,1}\,(\hat{y})$ which is the strongest near the intersection of`$u$' and `$\overline{B}$'. This nonuniform distribution of $\mathbf{j}_{diff,1}$ generates $\overline{\mathbf{b}}_{diff,\,1}$ along $\overline{U}$. This consequential process forms a net magnetic field $\overline{\mathbf{b}}_{net}=\overline{b}_{loc}\hat{x}+(\overline{b}_{nl}+\overline{b}_{diff,\,1})\hat{z}$ to be used as a new seed field for the next dynamo process. The outgrowing magnetic field along the velocity field $(\sim\hat{z})$ results in $\mathbf{b}_{net}$ closer to the velocity field. This geometry decreases EMF itself reducing dynamo efficiency. The conventional $\beta$ effect is based on this secondary interaction and represented as follows:
\begin{eqnarray}
-\int^t\langle \varepsilon_{ijk} u_j(t)u_l(\tau)\partial_l\overline{B}_k\rangle\, d\tau
&\rightarrow& -\frac{1}{3}\int^td\tau\langle u^2\rangle\nabla\times\overline{\mathbf{B}}\nonumber\\
&=&-\beta(t)\,\nabla\times\overline{\mathbf{B}}.
\label{beta_derivation_nonhelical}
\end{eqnarray}
The $\beta$ coefficient is always positive and diffuses magnetic energy to make the system homogeneous.\\

Fig.~2(b) and 2(c) show the evolution of a system forced with the left handed kinetic helicity\footnote{The left handed kinetic helicity ($f_h=-1$) is used for the visual simplicity in the plot. There is no practical difference from the right handed helical dynamo.}. The structures show  the poloidal velocity field $u_{pol}\,\hat{y}$ interacts with $\overline{b}_{nl}\,\hat{z}$ inducing the current density ${\bf u}_{pol}\times {\bf b}_{nl}\sim\overline{\bf j}_{ind,\,1}$ which is parallel to $\overline{\mathbf{B}}_{pol}$. Then, $\overline{\mathbf{j}}_{ind,\,1}$ generates the toroidal magnetic field $\overline{\mathbf{B}}_{tor}$ forming the right handed magnetic helicity with $\overline{\mathbf{B}}_{pol}$. These toroidal and poloidal magnetic field amplify each other through the $\alpha^2$ dynamo process so the strength and scale of this magnetic structure grow to surpass those of `other $\mathbf{b}$' fields. As $\overline{B}$ grows to be larger than other magnetic eddies $b$, the direction of $\overline{\mathbf{b}}_{loc}\,(\sim -{\bf u}\cdot \nabla \overline{\bf B})$ gets reversed from $\hat{x}$ to $-\hat{x}$ ($\nabla \overline{\bf B}<0 \rightarrow \nabla \overline{\bf B}>0$). Now, the magnetic energy in $\overline{\mathbf{B}}$ cascades toward $\mathbf{b}$ through this local transfer term.  Also, $\mathbf{u}_{pol}$ and $\overline{\mathbf{B}}_{pol}$ can interact with each other to yield the current density antiparallel to ${\bf b}_{nl}$, which can generate the left handed magnetic helicity. However, compared to $\mathbf{u}_{pol}\times \overline{\mathbf{b}}_{nl}$ this process is negligibly weak \citep{2019ApJ...872..132P}.\\

The growth of $\overline{B}_{pol}$ does not change the essential property of $\mathbf{u}_{pol}\times \overline{\mathbf{b}}_{nl}\sim \overline{\mathbf{j}}_{ind,\,1}$, but $\overline{\mathbf{j}}_{diff}$ is influenced by the relative strength of $\overline{B}_{pol}$. Referring to Fig.~2(b), 2(c), we may be able to expand $\overline{\mathbf{j}}_{diff}\sim\mathbf{u}(t)\times \int^t (-\mathbf{u}\cdot \nabla \overline{\mathbf{B}})\, d\tau$' as follows. Ignoring the integral symbol for simplicity, the current density is
\begin{eqnarray}
\overline{\mathbf{j}}_{diff}&\sim&-\xi_{ijk}u_j\langle (r,\,t)u_m(r+l,\,\tau)\rangle\frac{\partial \overline{B}_k}{\partial r_m}\label{beta_derivation_helical_first}\\
&\sim&-\xi_{ijk}\langle u_j(t)u_m(\tau)\rangle\frac{\partial \overline{B}_k}{\partial r_m}-\langle u_j(t)\,l_n\partial_n u_m(\tau)\rangle\xi_{ijk}\frac{\partial \overline{B}_k}{\partial r_m}\nonumber\\
&\sim&-\frac{1}{3}\langle u^2\rangle\xi_{ijk}\frac{\partial \overline{B}_k}{\partial r_j}-\xi_{jnm}\frac{l}{6}|H_V|\xi_{ijk}\frac{\partial \overline{B}_k}{\partial r_m}\delta_{nk}\delta_{mi}\nonumber\\
&\Rightarrow&\underbrace{-\frac{1}{3}\langle u^2\rangle \nabla\times \overline{\bf B}}_{1}+\underbrace{\frac{l}{6} |H_V|\nabla \times \overline{\bf B}}_{2}
\label{beta_derivation_helical1}
\end{eqnarray}
(We used $\langle u_j\partial_nu_m\rangle = \xi_{jnm}|H_V|/6$, where  $H_V$ is  $\langle \mathbf{u}\cdot\omega\rangle$. The subindices $m$, $n$ were chosen for the nontrivial result in the plots. Also, we assumed $l_n\rightarrow l$, but its physical meaning is not clear at present.)\\

With kinetic helicity, the effective $\beta$ coefficient can be represented as $\int^t\big(1/3\langle u^2 \rangle - l/6|H_V|\big)\,d\tau$. This indicates that any sign of kinetic helicity can amplify the large scale field as shown in Fig.~\ref{f1b}, \ref{f2b} (also refer to Eq.~(\ref{HmSolutionwithAlphaBeta1}), (\ref{EmSolutionwithAlphaBeta2})). This analytical approach explains the origin of $\beta$ in the nonhelical and helical case, but we need to analyze the mutual interactions of $U$ \& $B$ in the field structure for more detailed understanding.

\subsubsection{Kinematic Regime}
For $\partial \overline{B}_x/\partial z<0$ (Fig.~2(b)), the first term in Eq.~(\ref{beta_derivation_helical1}) can be written as $\overline{\mathbf{j}}_{diff,1}\sim1/3\langle u^2\rangle|\partial_z \overline{B}_x|\,\hat{y}$ leading to the increase of $\overline{\bf b}_{nl}(\hat{z})$. In contrast, the second term $\overline{\bf j}_{diff,\,2}\sim-l/6|H_V\partial \overline{B}_x/\partial z|\hat{y}$ decreases $\overline{\bf b}_{nl}$. The growing $\overline{\bf b}_{nl}$ increases $\overline{\bf j}_{ind,\,1}(\sim {\bf u}_{pol} \times \overline{\bf b}_{nl})$ and elevates the helical dynamo efficiency. The field analysis makes a consistent result but somewhat different representation. $\overline{\mathbf{j}}_{diff,\,1}\sim {\bf u}\times {\bf b}_{loc}$ is toward `$\hat{y}$' to increase $\overline{\bf b}_{nl}$. However, $\overline{\mathbf{j}}_{diff,\,2}\sim {\bf u}_{pol}\times {\bf b}_{loc}$ heads for `$-\hat{z}$' generating the left handed magnetic helicity with ${\bf b}_{nl}$, which reduces the overall dynamo efficiency. The difference between these two approaches actually comes from the fact that ${\bf u}_{pol}$ is included in $H_V$ and calculated separately from $\overline{B}$ (see Eq.~(\ref{beta_derivation_helical1})).\\


\subsubsection{Nonlinear Regime}
For $\partial \overline{B}_x/\partial z>0$ (Fig.~2(c)), the first term becomes $\overline{\mathbf{j}}_{diff,1}\sim-1/3\langle u^2\rangle|\partial_z \overline{B}_x|\,\hat{y}$. But, the second term is  $\overline{\bf j}_{diff,\,2}\sim l/6|H_V\partial \overline{B}_x/\partial z|\hat{y}$. $\overline{\mathbf{j}}_{diff,\,1}$ decreases $\overline{\bf b}_{nl}$ and helical dynamo efficiency ($\beta>0$). In contrast, the second term increases $\overline{\bf b}_{nl}$ and helical dynamo efficiency ($\beta<0$). The role of $\beta$ gets reversed compared to that of $\partial \overline{B}_x/\partial z<0$ in the kinematic regime. The field analysis also shows that the mutual interaction of ${\bf u}\times {\bf b}_{loc}$ yields $\overline{\mathbf{j}}_{diff,\,1}$ heading for `$-\hat{y}$'. On the contrary, ${\bf u}_{pol}\times {\bf b}_{loc}$ induces $\overline{\mathbf{j}}_{diff,\,2}$ parallel to $\overline{\bf b}_{nl}$ generating the right handed magnetic helicity.\\


On the other hand, the extraordinary change of $\beta$ in Fig.~\ref{f3a} can be explained with a virtual poloidal velocity field. The nonhelical forcing of the right handed helical field system can be realized as applying the left handed kinetic helicity to the system. We can assume a new poloidal velocity field $-u_{pol}\hat{y}$ with the same ${\bf u}_{tor}$ in the system. This new poloidal field can interact with $-{\bf u}\cdot \nabla {\bf b}$ to generate $-{\bf j}_{diff,\,2}\,(\hat{z})$ producing the left handed magnetic heicity, i.e., $\beta>0$.

\section{Summary}
We have discussed the physical meaning of $\alpha$ \& $\beta$ effect and how to find the coefficients using $\overline{E}_M(t)$ and $\overline{H}_M(t)$. We showed how these effects evolve with the large scale magnetic energy in the helical and nonhelical forcing dynamo. Fig.~1 indicates that the negative $\beta$ effect is a de facto dynamo generator after the $\alpha$ effect is quenched. To explain the results that are contradictory to the conventional dynamo theory, we used the field structure model and analytic method. According to this intuitive and analytical model, the $\beta$ effect is not fixed but evolves coupled with the relative strength of large scale magnetic field. For $-{\bf u}\cdot\nabla \overline{B}>0$ in the kinematic regime, $\overline{\mathbf{j}}_{diff,\,1}$ amplifies magnetic field ($\beta<0$), but $\overline{\mathbf{j}}_{diff,\,2}$ suppresses the growth of magnetic field ($\beta>0$). In contrast, for $-{\bf u}\cdot\nabla \overline{B}<0$ (nonlinear regime) $\overline{\mathbf{j}}_{diff,\,1}$ reduces the dynamo efficiency ($\beta>0$), but $\overline{\mathbf{j}}_{diff,2}$ elevates the growth of magnetic field ($\beta<0$). This kind of analysis may be inconsistent with the assumption of isotropy and homogeneous. However, while the helical system is isotropic and homogeneous macroscopically, the system is inhomogeneous and anisotropic without reflection symmetry microscopically. Also, helicity is a (peudo) scholar whose magnitude is arithmetically summed or detracted.\\ 

We may be tempted to ignore the $\alpha$ effect in dynamo. However, for the negative $\beta$ to become a helical dynamo generator, the nontrivial $\alpha$ effect that amplifies the large scale magnetic field beyond the kinematic regime is required. Moreover, without the $\alpha$ effect the vector potential `$A$' and magnetic field `$B_{tor}$' are just irrelevant fields (see  Eq.~(\ref{Solar_poloidal_magnetic_field}), (\ref{Solar_toroidal_magnetic_field})).







\bibliographystyle{mnras}
\bibliography{bibfile} 





\bsp	
\label{lastpage}
\end{document}